Deep Hedging, Generative Adversarial Networks, and Beyond

Hyunsu Kim

Northwestern University

**Abstract**

This paper introduces a potential application of deep learning / artificial intelligence in finance – particularly its application in hedging.

The major goal encompasses two objectives. First, we present a framework of a direct policy search reinforcement agent replicating a simple vanilla European call option and use the agent for the "model-free" delta hedging. Through this first objective, we would like to demonstrate that the neural nets as a universal approximation function – in conjunction with reinforcement learning - can play a role in solving advanced mathematical problems such as stochastic differential equation (SDE) / partial differential equation (PDE) in practice. In doing so, this will show how recurrent neural network (RNN)-based RL agents directly learn an optimal hedging strategy in a rather efficient manner given the risk aversion level of the investor - minimizing a convex risk measure like CVaR (Conditional Value at Risk) and producing better results particularly in terms of reducing tail risk exposures at higher values of the risk aversion parameter (i.e., 0.99). Our results (figure 10) clearly show that with the risk aversion parameter value of 0.99, all three RNN-based RL agents perform a lot better at minimizing the left tail exposures (i.e., minimizing the loss side on the PnL distributions) while naively embedding the additional risk aversion "constraint" in their learning processes.

Through the second part, in an attempt to alleviate the low-dimensional parametric process constraint - in doing so potentially bridging the gap between P-world and Q-world, we will illustrate nonparametric stochastic underlying scenarios generated by time-series GANs from multi-variate temporal space and demonstrate attempted hedging results on various values of risk aversion parameters. Our results (figures 20-21) show that with relatively higher excess kurtosis of PnL distributions (due to the very nature of fat tail empirically observed in the real markets), investors can utilize the risk-return trade-off – the more risk-averse the investors are (higher values of the risk aversion parameter), the lower the risk and returns are, vice versa – to meet a rather wide range of investment objectives while potentially achieving higher profits, in part, largely thanks to the nature of multi-variate temporal prediction embedded in nonparametric scenario generator.

    We believe that this RL-based hedging framework is a more efficient way of performing hedging in practice, addressing some of the aforementioned issues with the classic models, providing promising/intuitive hedging results, and rendering a flexible framework that can be easily paired with other AI-based models (i.e., GANs or VAE-GANs) for many other purposes (i.e., signal processing in high dimensional space and real-world market path generator).

## Introduction

In quantitative finance, traditionally two separate areas require quite advanced quantitative/mathematical techniques. As briefly mentioned by Attilio Meucci [1], on the one hand, the "Q" area of derivatives pricing deals with "extrapolating the present" by typically using Ito's calculus and partial differential equations (PDEs) under some risk-neutral probability measures (Q). On the other hand, the "P' area of quantitative risk/portfolio management deals with "modeling the future" by using multi-variate statistics and econometrics under some real-world probability measure (P).

Although those two areas seem to be distinctive and unique on each own, essentially two different sets of probabilistic weighting schemes are associated with the same possible outcomes given the same set of independent variables – understanding of risk premia makes a difference between them. For instance, the very gist of hedging is eventually protecting the future Profit&Loss (PnL) of a given position from a set of risk factors. This sounds like a concept of the P-world. But, in calculating Greeks (as an example, delta – option's sensitivity with respect to the underlying security), we classically use pricing models from Q-world, such as the Black-Scholes framework.

With that, the basic notion of performing hedging is simulating the dynamics of a financial market containing financial derivatives and relevant instruments and hedging the options in an investment portfolio by buying/selling the right amount of underlying stocks – therefore, eliminating risks. The hedging then, in a nutshell, can be considered as an algorithmic process where we re-adjust our hedge by rebalancing the number of shares we have in the underlying stock to match the new delta as the delta changes every period. This very concept, later known as delta hedging, was adopted by many of the prominent investment banks and hedge funds for years to come.

Such classic models as briefly aforementioned Black-Scholes, though, impose the assumptions on the dynamics of the underlying (stock movements) that are rather rigid with some parametric processes. Also, they are modeled in a low dimensional space with the underlying and a few other factors (i.e., volatility). As a result, this can, in turn, lead to a strong parametric process dependency concerning pricing/hedging parameters with a given classic model in a low dimensional space without:

- Naively/more efficiently dealing with other additional considerations such as investors' risk appetite and market frictions (i.e., transaction costs, liquidity constraints, market impacts, etc.)

- Fully utilizing additional state information available

Through the first part of this paper, we will demonstrate how the RNN-based direct policy search RL agents can perform delta hedging better than the classic Black-Scholes model (baseline) in Q-world based on parametrically generated underlying scenarios (in specific, by Geometric Brownian Motion), particularly minimizing tail exposures at higher values of the risk aversion parameter. In the second part of this paper, with the non-parametric paths generated by time-series GANs from multi-variate temporal space, we will also illustrate its delta hedging performance on various values of the risk aversion parameter via the basic RNN-based RL agent introduced in the first part of the research.

## Literature Review

Much of the related works on this front have been pioneered by many quantitative researchers with significant domain knowledge. A particular research work that is rather inspirational is the Deep Hedging paper written by Hans Buehler et al. back in 2018 [2]. In their paper, multiple sections detail a substantial amount of works on this front ranging from the overall set-up of a given problem (hedging) to theoretical/mathematical justifications behind using neural nets in this context.

In a nutshell, what Deep Hedging tries to achieve is given market signal information up to time t, liabilities, and a number of hedging instruments in a pre-defined asset universe, it tries to determine the most optimized holdings across different hedging instruments while meeting some of the key constraints (i.e., liquidity limit, trading costs, etc.) within their risk appetite (i.e., convex risk measure such as CVaR). And, this is done through some form of deep learning frameworks by providing "proper" rewards within a given environment for certain actions (how to allocate holdings across different assets to hedge liabilities) that the agent takes.

Without other complicated/additional concepts like the utility function overlay for the objective function construction, the overall framework is the following – discrete-time and market with friction assumed:

$$objective\ function\ of\ \pi(-Z) := \inf_{\delta \in H} \rho(-Z + P_0 + (\delta * S)_T - C_T(\delta)) \quad Eq.\ 1$$

$,where\ \mathbf{inf}\ indicates\ the\ greatest\ lower\ bound$

$,where\ \{I_0, \ldots, I_k\}\ is\ a\ set\ of\ market\ signals\ up\ to\ time\ k\ (t_k)\ that\ forms\ the\ filtration\ up\ to\ t_k$

$Z\ is\ a\ F_T\ measurable\ random\ variable\ indicating\ liabilities\ (or, contingent\ claims)$

$\delta_k^i\ is\ i^{th}\ asset\ holdings\ at\ the\ time\ t_k$

$dd\ H_k\ is\ a\ set\ of\ constraints\ that\ \delta_k\ is\ subject\ to\ at\ t_k$

$$(\delta * S)_T \text{ is defined as } \sum \delta_k * (S_{k+1} - S_k)$$

$$C_T(\delta) \text{ is defined as } \sum C_k(\delta_{k+1} - \delta_k)$$

, where $C_k$ can take fixed, proportional, and rather complicated cross

− asset cost functional forms

$P_0$ is defined as cash injection or extraction

, where $\delta_k^\theta := f^\theta(I_k, \delta_{k-1}^\theta)$

$\theta$ indicates a set of parameters for the trained neural net

$f$ is a composite functional form of neural nets $\left(i.e. f\left(g\left(k(h(x))\right)\right)\right)$

$\delta_{t-1}^\theta$ indicates the recurrent nature of the neural nets

$\rho$ is a convex risk measure meeting the three following properties:

− Monotone decreasing: if $x_1 \geq x_2$, then $\rho(x_1) \leq \rho(x_2)$

; in words, this means more favroable positions require less cash injection

− Convex: $\rho(\alpha x_1 + (1 - \alpha)x_2) \leq \alpha\rho(x_1) + (1 - \alpha)\rho(x_2)$

; in words, diversification works

− Cash − invariant: $\rho(x + c) = \rho(x) - c$, where $c \in \mathbb{R}$

; in words, adding cash to a position reduces the need for more by that amount

In a nutshell, what it is after using the objective function here is getting the most optimized holdings across different hedging instruments while meeting some of the key constraints (i.e., liquidity limit, trading costs, etc.) within their risk appetite (i.e., convex risk measure such as CVaR).

The first part of our paper is directly inspired by the Deep Hedging framework introduced by Hans Buehler et al., particularly focusing on delta hedging performance of Black-Scholes and RNN-based RL agents in the presence of the risk aversion parameter – without additional implications such as transaction costs, liquidity constraints, market impacts, etc.

However, in the second part of our research, we further extend this idea by bringing in non-parametric underlying stochastic scenarios generated by time-series generative adversarial networks (GANs) from multi-variate temporal space. A particularly elegant framework in dealing with multi-variate temporal space was introduced by Jinsung Yoon et al. [3] with novel time-series GANs and it articulated how the framework can be used to generate synthetic data given historical multi-variate information available. This led us to believe that this general framework (or, beyond this framework such as time-series VAE-GAN expressed in rather continuous latent space with the Bayesian framework overlay) can be served as:

- A rather realistic time-series "simulator" generating potential market paths from multi-variate temporal space without imposing parametric process constraints
- A multi-variate time-series forecasting model taking into account both temporal dependencies and dependency structures across different state-space variables

With the non-parametric scenarios generated by this framework, through the second part of this paper, we will demonstrate the delta hedging performance of a basic RNN-based RL agent introduced in the first part across various values of the risk aversion parameter.

## Part I – "Model-free" RL Agents for Delta Hedging in Q-world

<u>Data</u>

Both training and test data sets will be generated by the Monte Carlo simulation followed by Geometric Brownian Motion (GBM). The basic notion is that the underlying (stock price) follows a random walk with a Markov process where it effectively indicates the past information is already incorporated and the next price movement is conditionally independent of the history of stock movements.

$$\Delta S = S * (\mu \Delta t + \sigma \Delta z)$$

$$, where\ S = current\ stock\ price\ (underlying)$$

$$\mu = the\ expected\ return$$

$$t = time\ to\ expiration\ of\ the\ option$$

$$\sigma = standard\ deviation\ of\ log\ return$$

$$\Delta z \sim N(0, \Delta t)$$

Note that in this first part of the research, we prepared two separate data sets for training and test purpose (in-sample / out-of-sample). Each dataset contains 500,000 generated scenarios across 30 different time steps. The number of features is set to 1 since we are dealing with only the underlying stock values.

<u>Method</u>

As opposed to other well-known/famous reinforcement learning methods such as off-policy Q-learning and Monte Carlo Tree Search (MCTS), we will be using a policy search method where an agent directly learns how to map a given state to a particular action without learning action values to determine action selections. Simply put, the direct policy search is defined as an RL method searching for the direction where a given policy network's parameters change towards improving the current policy – a more rewarding direction.

As detailed by Richard S. Sutton et al. [4], this direct policy search method can entail many advantages over action value-based methods. A couple of them are:

- Asymptotically converging to deterministic policies with right levels of exploration
- Natively handling continuous action spaces well
- Being better suited for some problems to parametrize the policy rather than the value/action-value functions

The overall framework of the direct policy search method is the following:

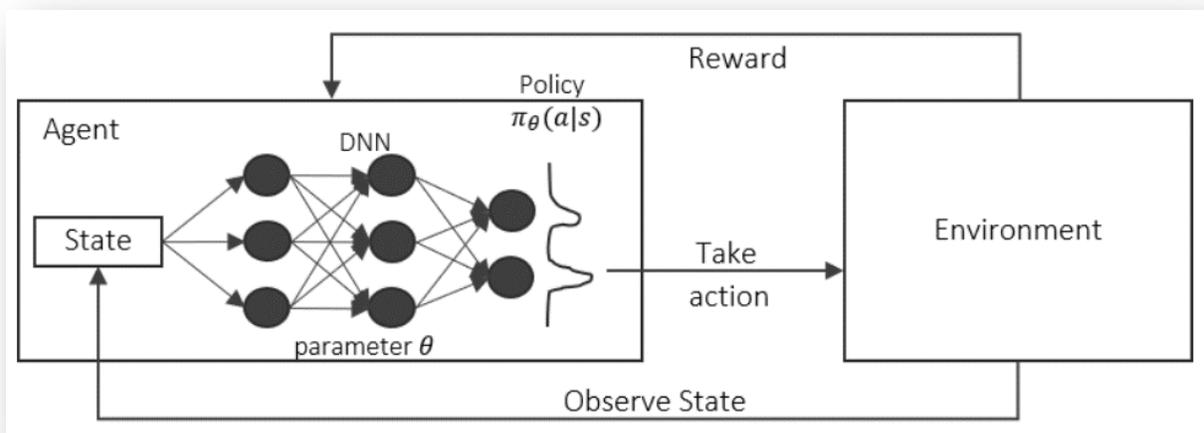

Figure 1: Generic Description of Direct Policy Search

- Policy: This refers to a probability of an action given state.

$$\pi_\theta(a|s) = P(a|s, \theta)$$

*, where a is action a in action space*

*s is state s in state space*

*θ is a set of parameters*

- Objective function: Intuitively (or, heuristically), the objective function is to maximize (or, minimize) the expected rewards (or, penalties) by taking optimized actions given states – optimal policy. And, the optimal policy can be set by finding a set of parameters, $\theta$ through DNN (Deep Neural Nets).

$$J(\theta) \propto \sum_{s} \mu(s) \sum_{a} \nabla \pi_\theta(a|s) q_\pi(s,a) \qquad Eq.\ 2$$

, where $q_\pi(s,a)$ is quality of state and action following the policy

$\pi_\theta(a|s)$ is probability of taking action a given state s

$\mu(s)$ is the on − policy distribution

(can be considered as weights on different state samples)

This framework intuitively makes sense as the gradient scale is proportional to quality (value) of action taken given state – since we would like to increase the probability of action taken that provides the higher value of that action given state, vice versa. The parameters are then updated through some form of stochastic gradient descent optimizer like the following:

$$\theta^* \leftarrow \theta + \alpha \nabla_\theta J(\theta) \qquad Eq.\ 3$$

, where $\alpha$ is a learning rate

$\nabla_\theta J(\theta)$ is a gradient with respect to a set of policy parameters

With this general framework of the direct policy search in mind, the detailed framework of our hedging model is the following:

$$objective\ function\ of\ \pi(-Z) := \inf_{\delta \in H} \rho(-Z + P_0 + (\delta * S)_T) \qquad Eq.\ 4$$

, where **inf** indicates the greatest lower bound

$, where \{S_0, ..., S_k\}$ is a set of market signals up to time $k$ $(t_k)$ that forms the filtration up to $t_k$

$Z$ is a $F_T$ measurable random variable indicating liabilities (or, contingent claims)

$\delta_k^i$ is ith asset holdings at the time $t_k$

$H_k$ is a set of constraints that $\delta_k$ is subject to at $t_k$

$(\delta * S)_T$ is defined as $\sum \delta_k * (S_{k+1} - S_k)$

$P_0$ is defined as cash injection or extraction

$\rho$ is a convex risk measure, CVaR

$, where\ \delta_k^\theta := f^\theta(I_k, \delta_{k-1}^\theta)$

$\theta$ indicates a set of parameters for the trained neural net

$f$ is a composite functional form of neural nets $\left(i.e. f\left(g\left(k(h(x))\right)\right)\right)$

$\delta_{k-1}^\theta$ indicates the recurrent nature of the neural nets

Then, connecting dots with the generic direct policy search, we can frame our framework as the following reinforcement learning problem:

- State:

    $\{S_0, ..., S_k\}$ is a set of underlying values up to time $k$ $(t_k)$

- Action:

    Univariate continuous action space of $\delta_k$ at each time $t_k$

- Reward:

$$CVaR(S, K, P_0, \delta_t, \alpha) = CVaR(-\max(S_T - K, 0) + \sum \delta_t * (S_{t+1} - S_t) + P_0, \alpha) \quad Eq.\ 5$$

$, where\ \alpha$ indicates risk aversion parameter (or, risk appetite)

### Experimental Setup

Through the first part of this paper, we will demonstrate how the RNN-based direct policy search RL agents can perform delta hedging for a simple vanilla European call option better than the classic Black-Scholes model (baseline) with parametrically generated stochastic scenarios as inputs, particularly minimizing left tail exposures with higher values of the risk aversion parameter.

Throughout the experiments in this first part of the research, the following table contains the initial parameter setting for the Black-Scholes framework and the shared hyper-parameter setting for each RNN-based agent:

| | |
|---|---|
| **Initial underlying value** | 100 |
| **Strike value** | 100 (at-the-money) |
| **Risk-free rate** | 0 |
| **Volatility** | 0.15 |
| **Time-to-maturity** | 1/12 (annual basis – one month) |
| **Time steps in given month** | 30 (30 calendar days) |
| **# of RNN nodes at each layer** | 128, 64, 64, 1 |
| **Training epochs** | 50 |
| **Batch size** | 1000 |
| **Alpha (risk aversion parameter)** | 0.5, 0.75, 0.99 |
| **Optimizer** | Adam |

Table 1: Initial Parameter Setting

## Results

The overall hedge PnL distributions and evolutions of deltas across three different values of the risk aversion parameter are illustrated in the following figures for Black-Scholes, basic RNN, LSTM, and GRU, respectively – all based on out-of-sample data.

- Risk aversion parameter ($\alpha$) = 0.5

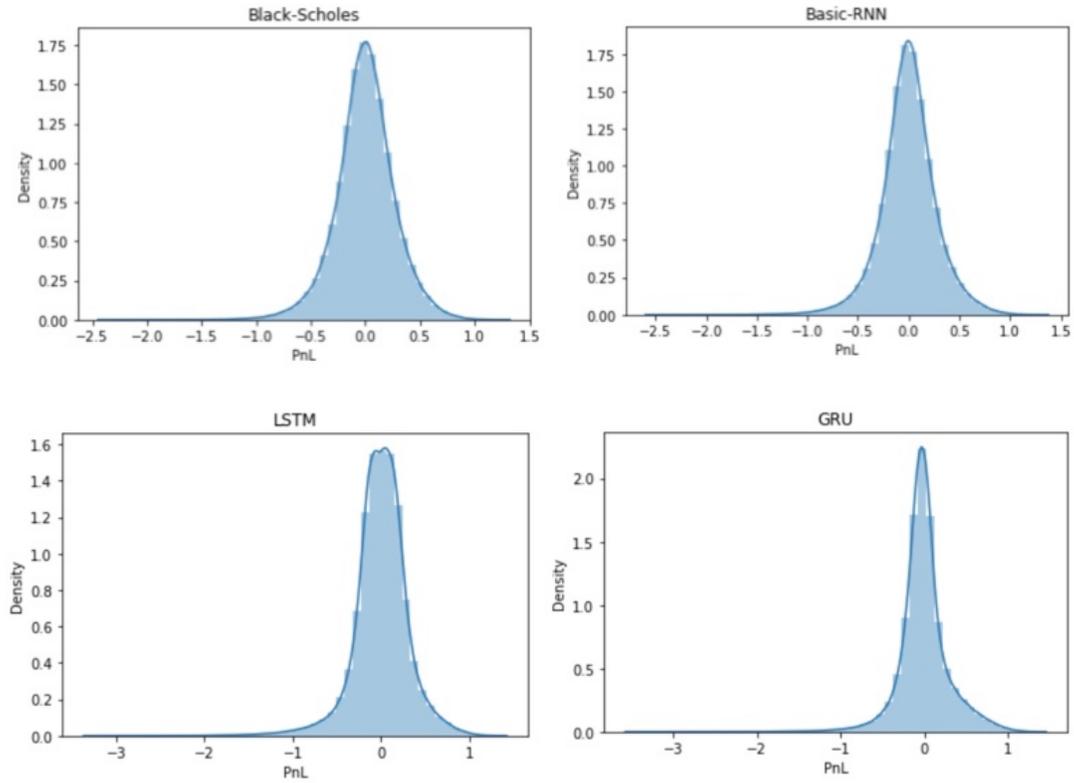

| $\alpha = 0.5$ | Black-Scholes | Basic-RNN | LSTM | GRU |
|---|---|---|---|---|
| Mean PnL | 0.0005 | 0.0006 | 0.0004 | 0.0007 |
| CVaR PnL ($\alpha = 0.5$) | -0.2028 | -0.2040 | -0.2197 | -0.2000 |
| Skewness | -0.2515 | -0.2922 | -0.7912 | -0.1628 |
| Excess Kurtosis | 1.5200 | 2.2705 | 4.4625 | 4.3622 |

Figure 3: Overall PnL Distributions and Key Stats of BS and RNN-based RL Agents at $\alpha = 0.5$

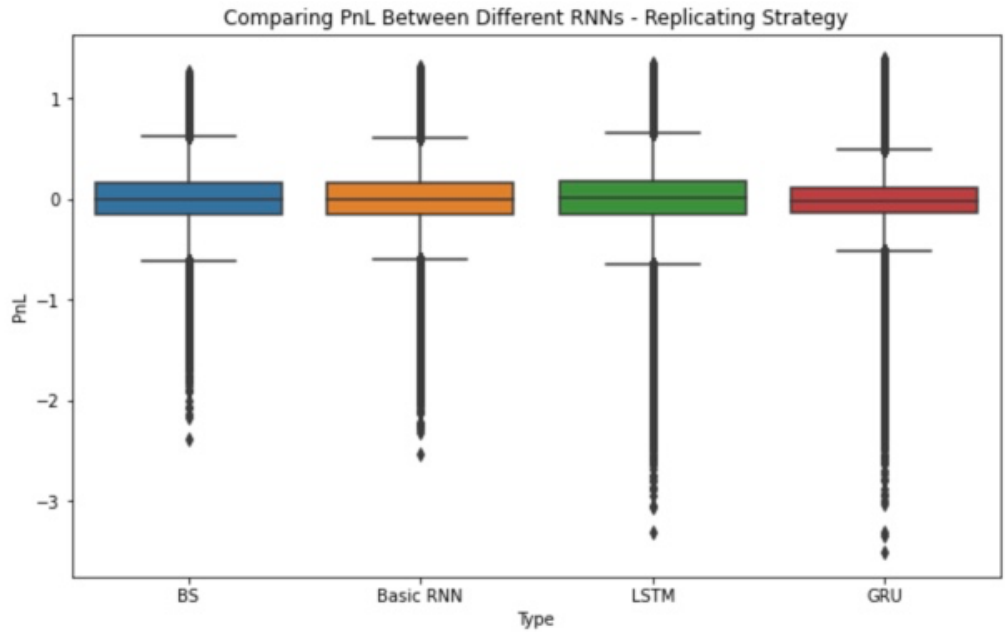

Figure 4: Boxplot Comparison for PnL Distributions at $\alpha = 0.5$

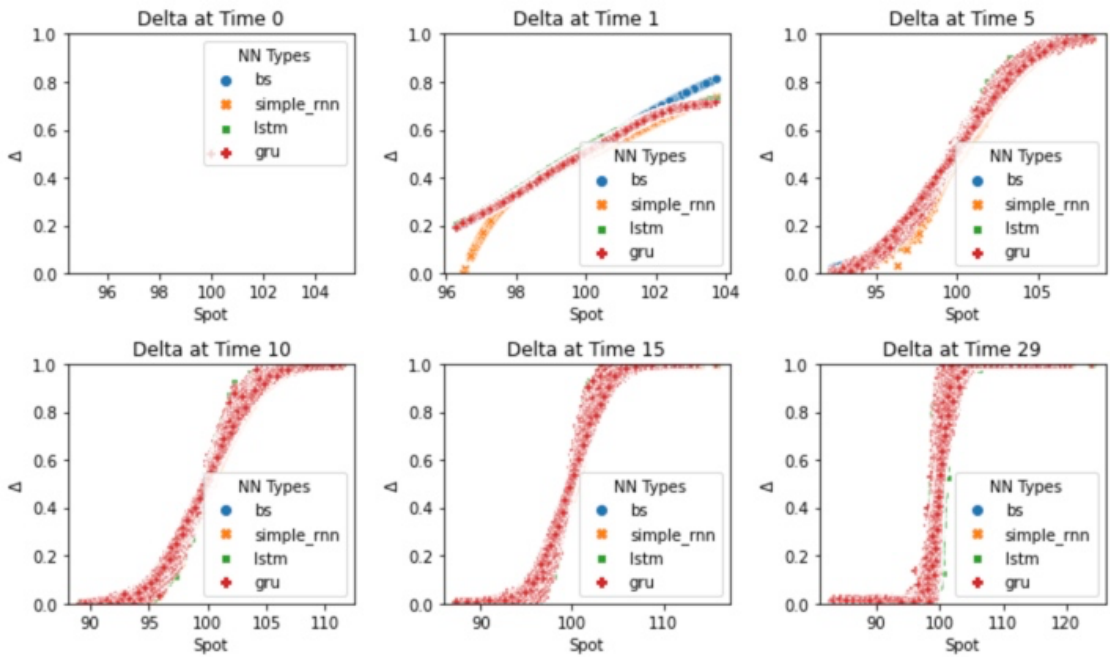

Figure 5: Evolution of Delta – Convexity Observations for Every Type at $\alpha = 0.5$

- Risk aversion parameter $(\alpha) = 0.75$

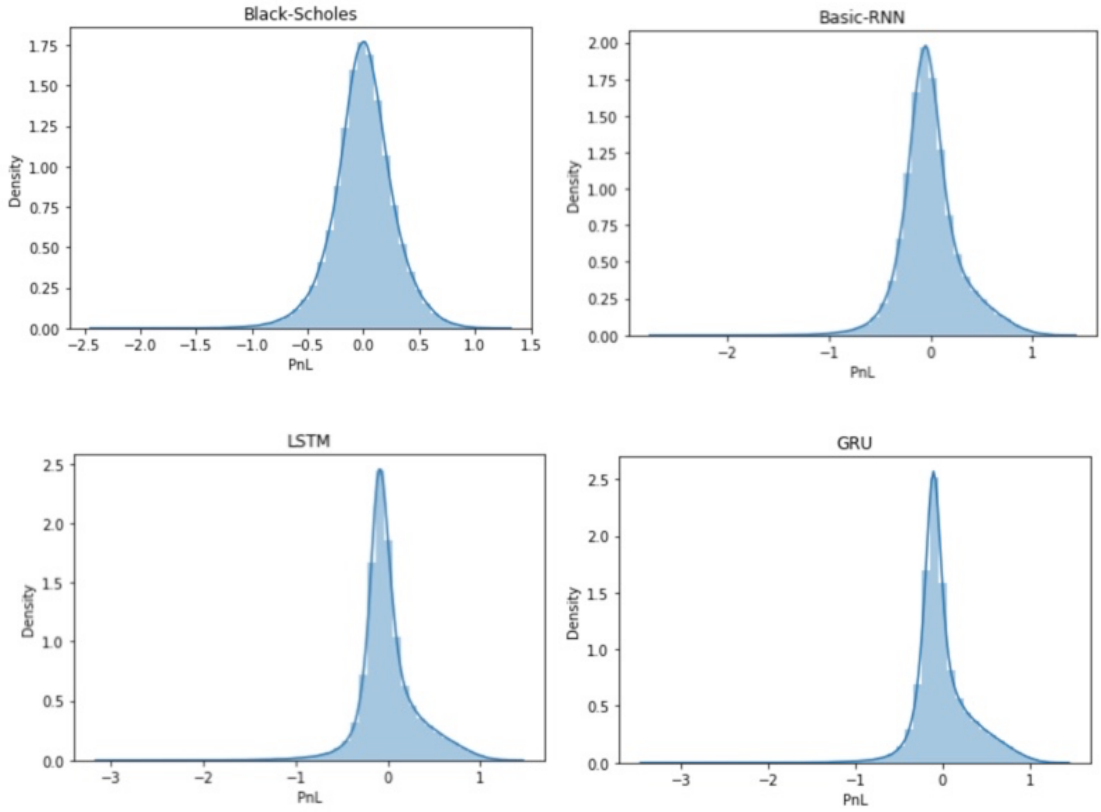

| $\alpha = 0.75$ | Black-Scholes | Basic-RNN | LSTM | GRU |
|---|---|---|---|---|
| Mean PnL | 0.0005 | 0.0005 | 0.0008 | 0.0007 |
| CVaR PnL ($\alpha = 0.75$) | -0.3353 | -0.3257 | -0.3132 | -0.3119 |
| Skewness | -0.2515 | 0.2798 | 0.2359 | 0.4428 |
| Excess Kurtosis | 1.5200 | 2.2931 | 3.7518 | 3.1937 |

Figure 6: Overall PnL Distributions of BS and RNN-based RL Agents at $\alpha = 0.75$

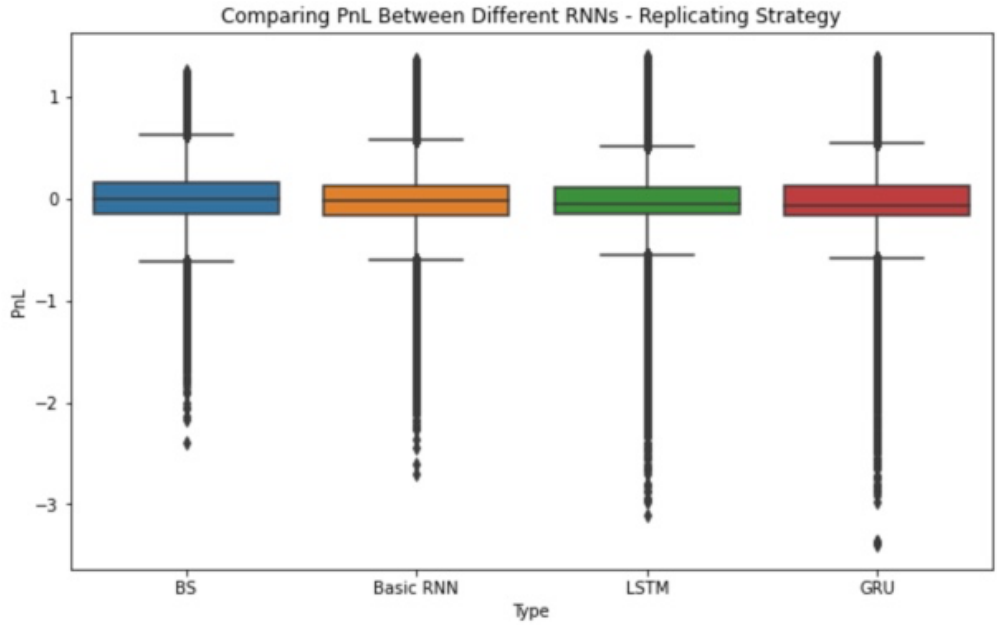

Figure 7: Boxplot Comparison for PnL Distributions at $\alpha = 0.75$

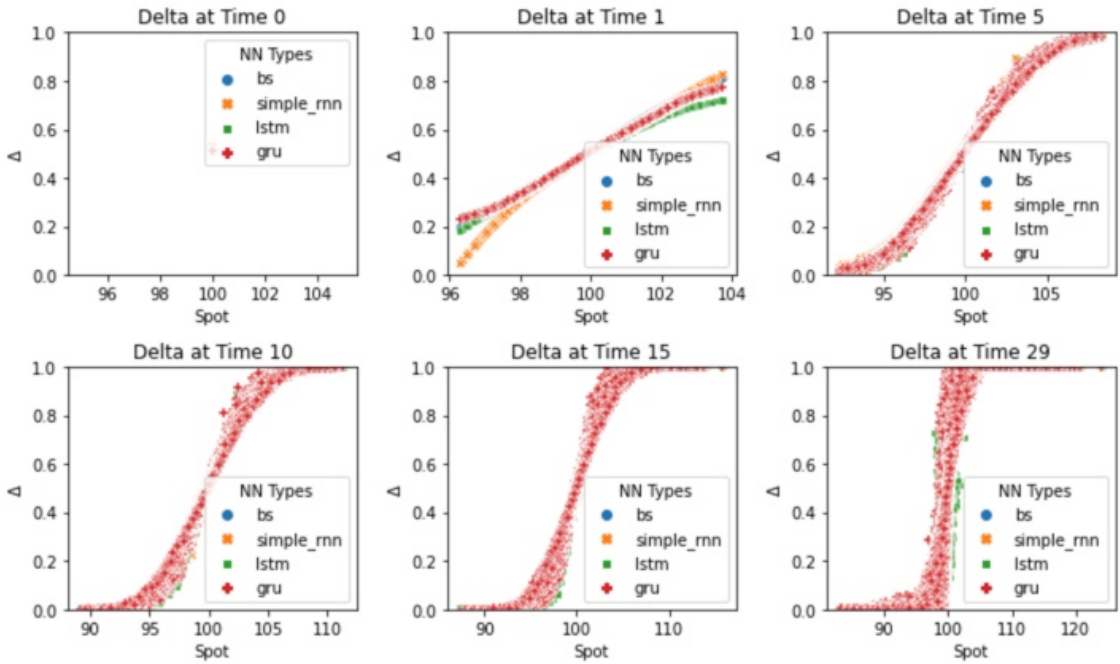

Figure 8: Evolution of Delta – Convexity Observations for Every Type at $\alpha = 0.75$

- Risk aversion parameter ($\alpha$) = 0.99

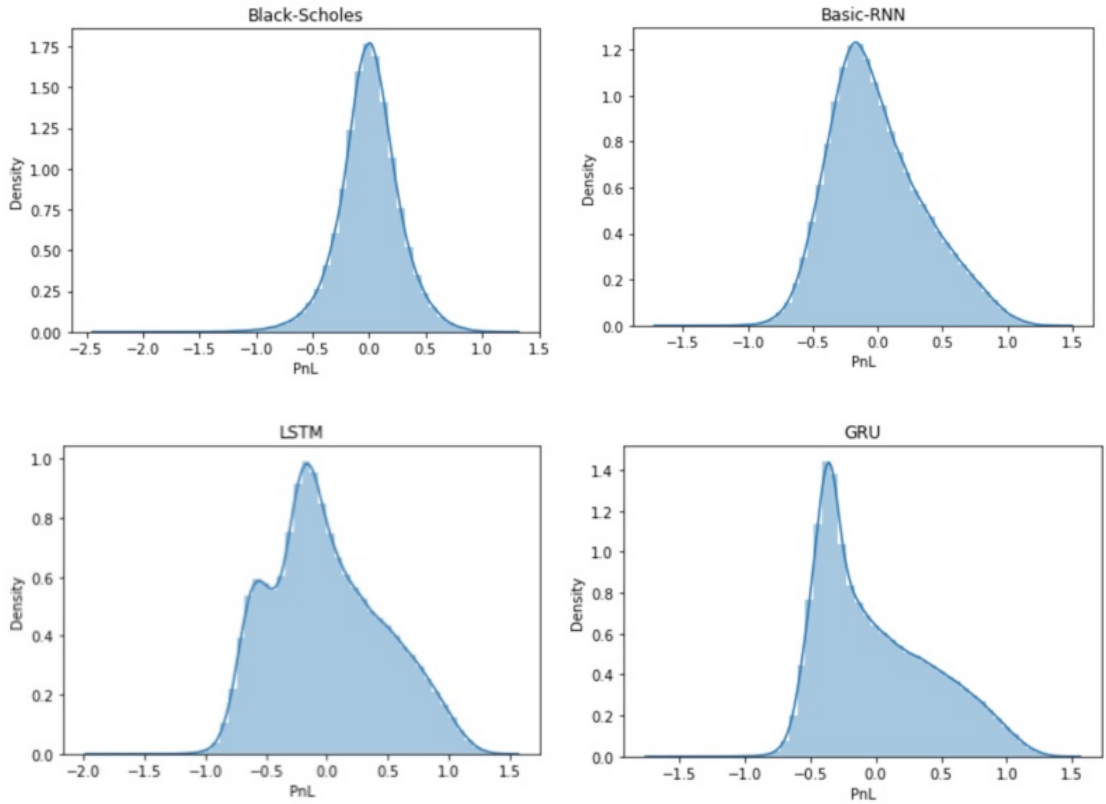

| $\alpha = 0.99$ | Black-Scholes | Basic-RNN | LSTM | GRU |
|---|---|---|---|---|
| Mean PnL | 0.0005 | 0.0005 | 0.0011 | 0.0007 |
| CVaR PnL ($\alpha = 0.99$) | -0.9203 | -0.7810 | -0.8974 | -0.7270 |
| Skewness | -0.2515 | 0.5276 | 0.3653 | 0.6776 |
| Excess Kurtosis | 1.5200 | -0.0553 | -0.5312 | -0.4607 |

Figure 9: Overall PnL Distributions of BS and RNN-based RL Agents at $\alpha = 0.99$

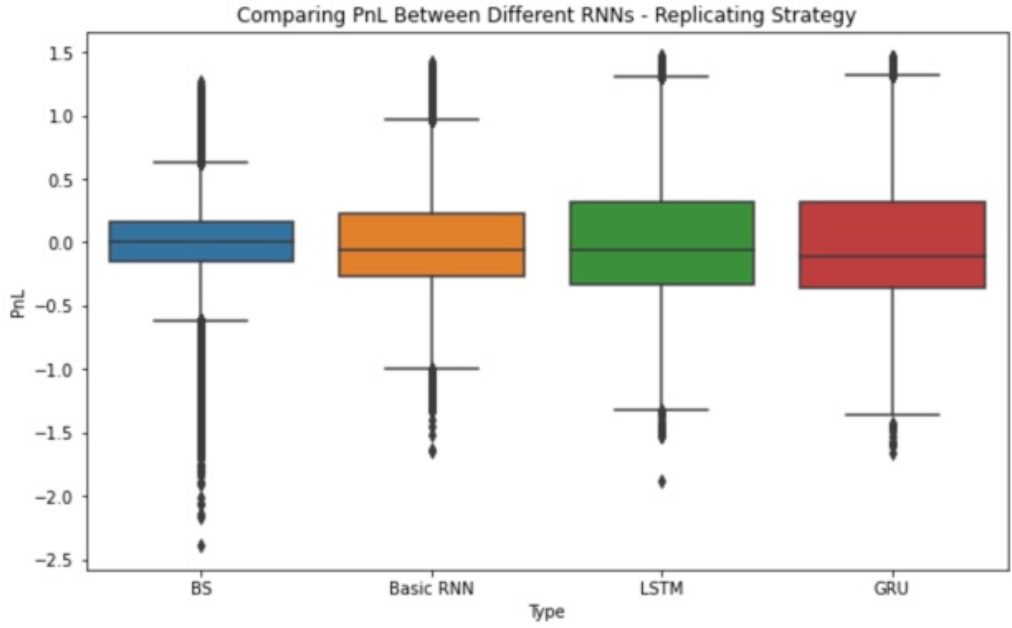

Figure 10: Boxplot Comparison for PnL Distributions at $\alpha = 0.99$

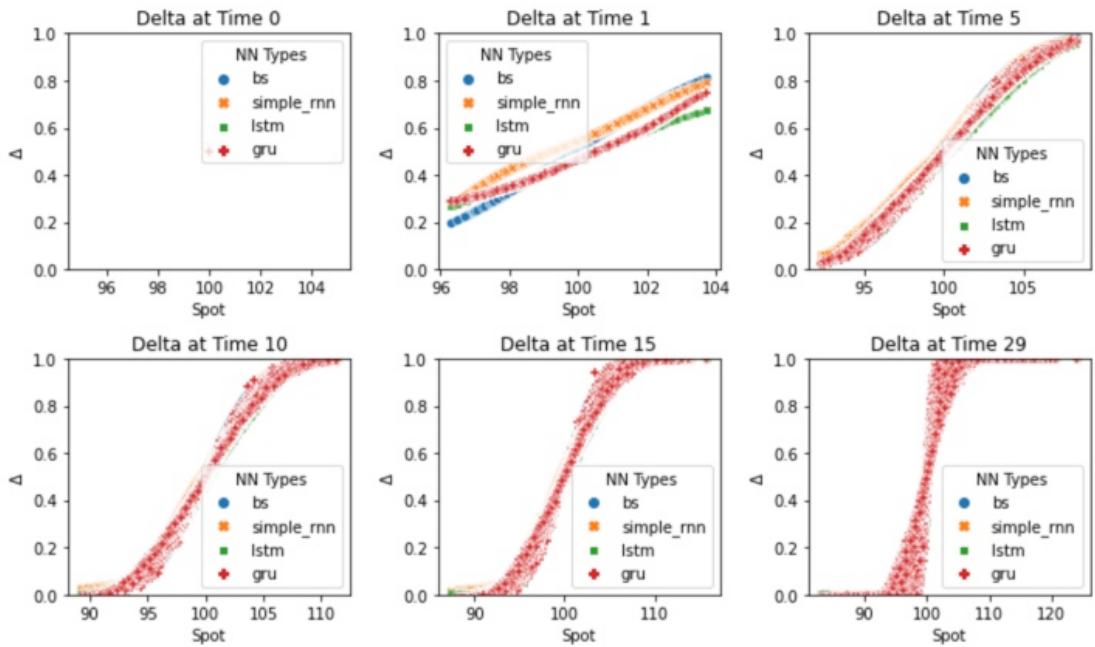

Figure 11: Evolution of Delta – Convexity Observations for Every Type at $\alpha = 0.99$

Interpretation

At both the risk aversion parameter ($\alpha$) values of 0.5 and 0.75 (figures 3-7), both mean and CVaR values of all three RNN-based RL agents quite closely matched the ones of Black-Scholes. In particular, the belly part (from the lower quartile to the upper quartile – figures 4 and 7) of PnL distributions were tightly distributed both for $\alpha$ of 0.5 and 0.75 while lowering excess kurtosis at $\alpha$ of 0.75 – indicating a bit of minimization on the left tail exposures as we increased the risk aversion parameter value.

At $\alpha$ of 0.99 (figures 9-10), mean values of all three RNN-based RL agents are still fairly close to the mean value of Black-Scholes (except for LSTM) whereas CVaR values are distinctively smaller than the CVaR value of Black-Scholes. This can be confirmed by the fact that the excess kurtosis values of all of the other three agents result in a lot lower values in comparison to Black-Scholes. The main reason would be that all of the three agents were natively trained with a loss function whose objective is to minimize the left tail exposure, namely the hedging loss. In specific, in our experiment with $\alpha$ of 0.99, the learning objective of three agents was to minimize the worst 1 in 100 hedge loss events – say 10 worst hedge losses out of 1000 events (definition of CVaR with $\alpha$ of 0.99). As a result, the overall distribution of PnLs from all of the three agents is expected to show rather thinner left tails compared to the Black-Scholes framework – since they were natively/efficiently learned to achieve that goal by directly finding parameterized policies. This, in turn, can be rather clearly seen in Figure 11 looking at the left tail side of PnL distributions – also, interestingly the three agents all showed fatter belly parts compared to Black-Scholes (from the lower quartile to the upper quartile).

Furthermore, we can confirm that the delta values of a call option were captured well across different time horizons with the classic convexity nature in mind – a stylized fact (figures 5, 8, and 11).

From a theoretical perspective, we demonstrated that the neural nets as a universal approximation function – in conjunction with reinforcement learning (direct policy search method in our case) - can play a role in solving advanced mathematical problems such as stochastic differential equation (SDE) / partial differential equation (PDE) in practice. From a practical perspective, our results showed how recurrent neural network (RNN)-based RL agents directly learn an optimal hedging strategy in a rather efficient manner given the risk aversion level of the investor - minimizing a convex risk measure like CVaR and producing better results particularly in terms of reducing left tail risk exposures at higher values of the risk aversion parameter (i.e., 0.99) while naively embedding the additional risk aversion "constraint" in their learning processes.

**Part II – "Model-free" RL Agents for Delta Hedging with Nonparametric Underlying Paths**

Data

Both training (1/3/1950 – 1/25/2010) and test data (1/3/1950 – 1/25/2021) sets will be generated by time-series GANs taking into account the multi-variate temporal aspects of the data. Each dataset contains 500,000 generated scenarios across 30 different time steps. However, the GANs will be trained in the multi-variate temporal space – time-series of six features (open, high, low, close, adjusted close, and volume). That way stochastic paths of adjusted close generated by the GANs natively stem from some form of a nonparametric joint distribution of the initial six multi-variate dimensions.

Method

The hedging methodology will follow the same one as the first part of the research above – a direct policy search RL. Now, what is different in this section is the input data generated by time-series GANs from multi-variate temporal space.

The GANs were inspired by the very notion of the Turning test (Turing learning) where an artificial intelligence agent keeps trying to beat a given system (critic). In doing so, the system is eventually becoming convinced that the agent is at a near-human level for achieving a given goal. With the high-level mathematical framework of the classic GANs below (figure 12) in mind, the optimization of GANs essentially keeps carrying out until it reaches the so-called "Nash equilibrium" where both parties in this game (generator and discriminator) have reached their peak abilities so that the optimization does not improve any further – which is somewhat opposite of the modal collapse.

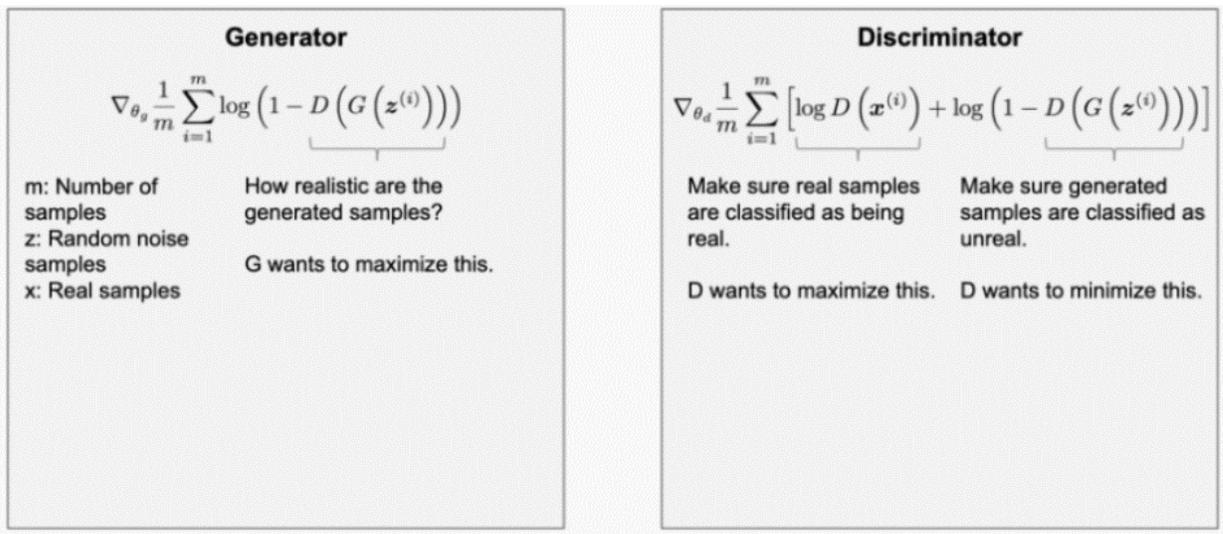

Figure 12. Generic Framework of GANs

On top of its very nature of difficult convergence, dealing with multi-variate time-series poses additional challenges in terms of properly constructing a generative model. The reason being is that the model not only has to capture the distributions of independent variables within each time step but also should be able to accurately capture the serial dynamics across different time steps.

The time-series GANs proposed by Jinsung Yoon et al. [2] took into account these aspects of consideration. By imposing the auto-encoder structure on top of the GANs, the adversarial nets can more efficiently play their game (between discriminator and generator) at a lower dimension latent space. Besides, it overlays a stepwise supervised loss based on the real data so that it facilitates the model to better preserve the stepwise conditional distributions in the data. And, minimizing this supervised loss jointly carries out with the reconstruction loss (from the auto-encoder part) and the unsupervised loss (from the adversarial networks). In doing so, the embedded latent space preserves both temporal and multi-variate dynamics concurrently in a lower-dimensional space. The following figure 13 shows the overall architecture of time-series GANs quite neatly.

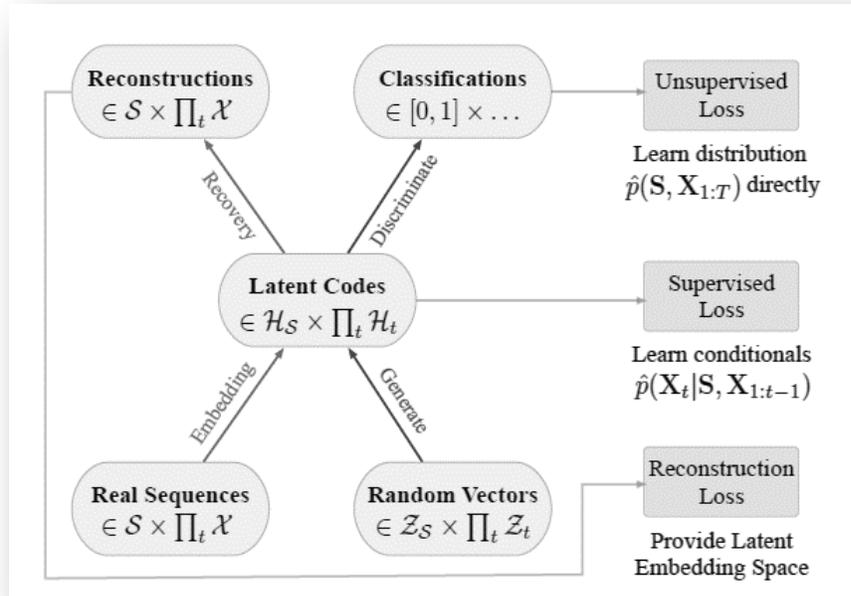

Figure 13. Building Blocks of Time-series GANs

Experimental Setup

In the second part, in an attempt to bridge the gap between P-world and Q-world, we will show nonparametric stochastic underlying scenarios generated by time-series GANs (real-world P-measure) from multi-variate temporal space and demonstrate the corresponding hedging results on various values of risk aversion parameters, hoping that with higher values of risk aversion parameters, we can still minimize the left tail exposures in terms of each RL's PnL distribution and show a rather clear risk-return trade-off observed in the real markets.

Throughout the experiments in this second part of the research, the following two tables contain the parameter settings for time-series GANs and a basic RNN-based RL agent:

| Sequence length | 31 |
|---|---|
| # of hidden nodes in each RNN layer | 31 |
| # of RNN layers | 3 |
| Iteration | 10,000 |
| Batch size | 178 |
| Optimizer | Adam |

Table 2: Initial Parameter Setting for Time-series GANs

| | |
|---|---|
| Initial underlying value | 100 |
| Strike value | 100 (at-the-money) |
| Risk-free rate | 0 |
| Volatility | 0.15 |
| Time-to-maturity | 1/12 (annual basis – one month) |
| Time steps in given month | 30 (30 calendar days) |
| # of RNN nodes at each layer | 128, 64, 64, 1 |
| Training epochs | 50 |
| Batch size | 1000 |
| Alpha (risk aversion parameter) | 0.5, 0.75, 0.99, 0.995, 0.997 |
| Optimizer | Adam |

Table 3: Initial Parameter Setting for Basic RNN-based RL Agent

## Results

Both for training and testing data, simple visualizations are illustrated in a 2-D low dimensional space by t-SNE. With those nonparametrically generated data, the overall hedge PnL distributions across three different values of the risk aversion parameter are shown in the following figures for a basic RNN-based RL agent.

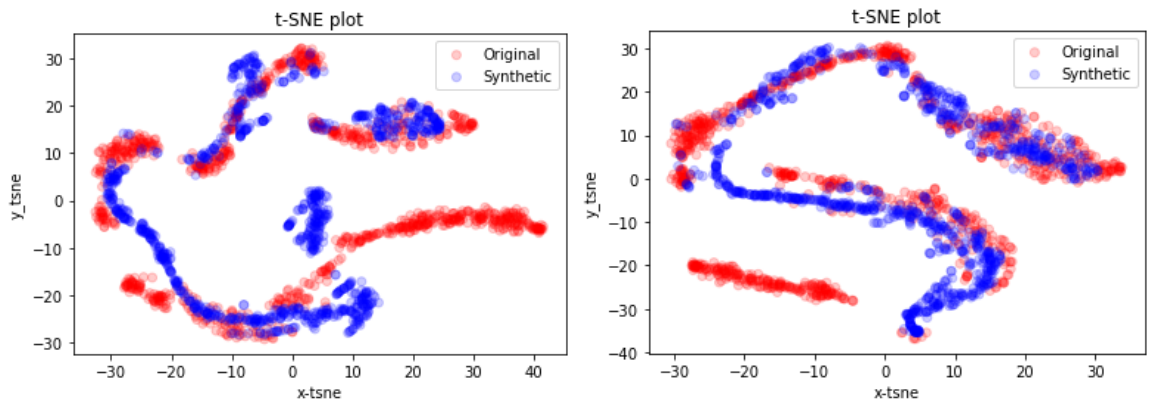

Figure 14. Real Data vs. Generated Data for both Training (Left) and Test (Right) Data

- Risk aversion parameter ($\alpha$) = 0.5

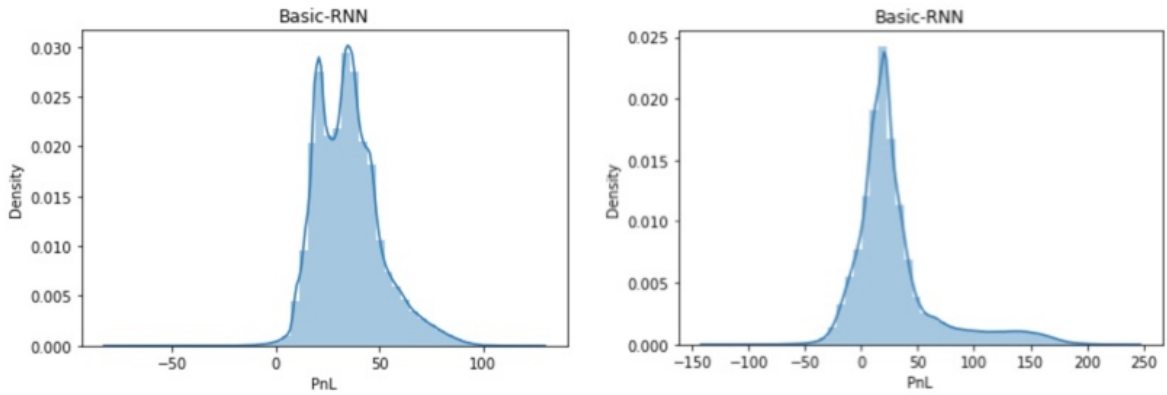

| $\alpha$ = 0.5 | In-Sample | Out-of-Sample |
|---|---|---|
| Mean PnL | 35.1156 | 29.7132472 |
| CVaR 1st PnL | 31.5402 | 19.08048912 |
| CVaR 25th PnL | 28.1654 | 12.70764771 |
| CVaR 50th PnL | 23.0647 | 5.496739995 |
| CVaR 75th PnL | 17.3686 | -4.20621819 |
| CVaR 99th PnL | 2.3901 | -36.9978025 |
| Skewness | 0.7530 | 1.789179899 |
| Excess Kurtosis | 0.8719 | 3.568665797 |

Figure 15. Overall PnL Distribution of Basic RNN for In-Sample (Left) & Out-Of-Sample (Right) at $\alpha$ = 0.50

- Risk aversion parameter ($\alpha$) = 0.75

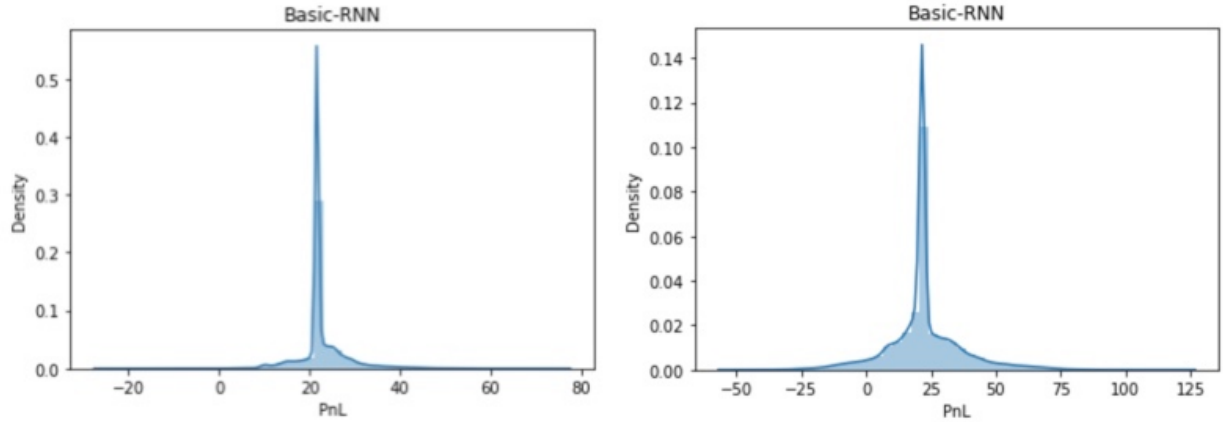

| $\alpha = 0.75$ | In-Sample | Out-of-Sample |
|---|---|---|
| Mean PnL | 22.6622 | 22.3680 |
| CVaR 1st PnL | 21.4015 | 19.2192 |
| CVaR 25th PnL | 20.5977 | 16.6774 |
| CVaR 50th PnL | 19.9495 | 13.7959 |
| CVaR 75th PnL | 18.2912 | 6.7378 |
| CVaR 99th PnL | 5.6282 | -22.1774 |
| Skewness | 1.2167 | 0.3533 |
| Excess Kurtosis | 7.4380 | 2.8329 |

Figure 16. Overall PnL Distribution of Basic RNN for In-Sample (Left) & Out-Of-Sample (Right) at $\alpha = 0.75$

- Risk aversion parameter ($\alpha$) = 0.99

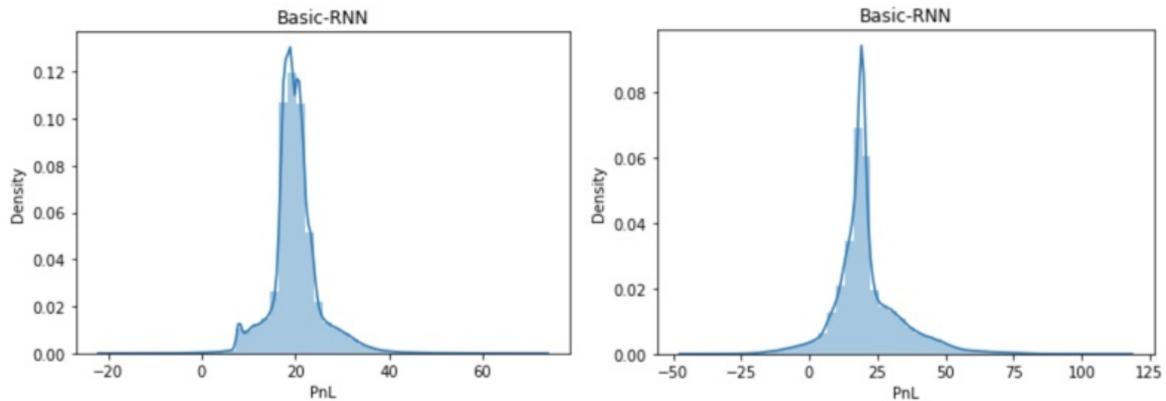

| $\alpha = 0.99$ | In-Sample | Out-of-Sample |
|---|---|---|
| Mean PnL | 19.9613 | 21.1204 |
| CVaR 1st PnL | 18.7896 | 18.2166 |
| CVaR 25th PnL | 17.8991 | 15.7924 |
| CVaR 50th PnL | 16.4989 | 13.1239 |
| CVaR 75th PnL | 14.4047 | 8.4799 |
| CVaR 99th PnL | 4.7712 | -13.7966 |
| Skewness | 0.6936 | 0.8800 |
| Kurtosis | 4.3637 | 2.9194 |

Figure 17. Overall PnL Distribution of Basic RNN for In-Sample (Left) & Out-Of-Sample (Right) at $\alpha = 0.99$

- Risk aversion parameter ($\alpha$) = 0.995

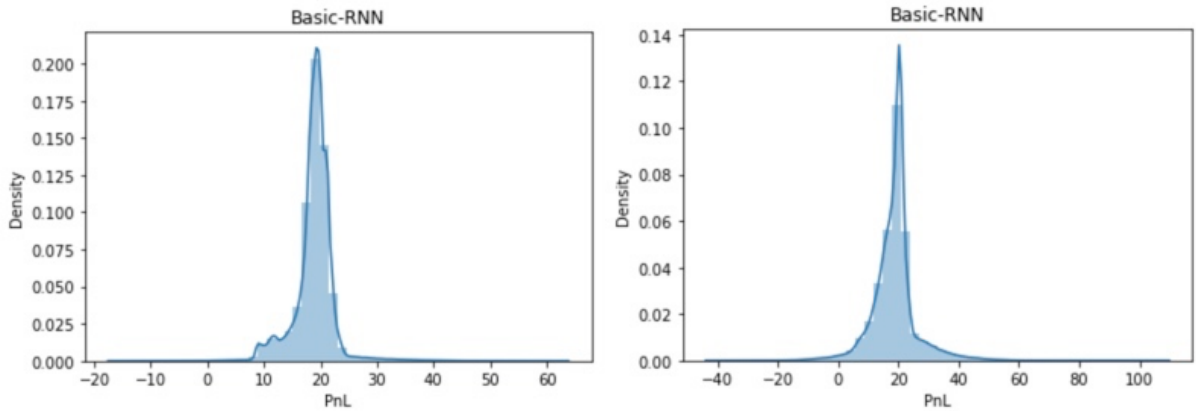

| $\alpha = 0.995$ | In-Sample | Out-of-Sample |
|---|---|---|
| Mean PnL | 18.8863 | 18.8445 |
| CVaR 1st PnL | 18.2697 | 17.2436 |
| CVaR 25th PnL | 17.7119 | 16.1910 |
| CVaR 50th PnL | 16.6629 | 14.1377 |
| CVaR 75th PnL | 14.7597 | 10.5980 |
| CVaR 99th PnL | 6.6959 | -7.3296 |
| Skewness | 0.3567 | 0.4759 |
| Excess Kurtosis | 9.5104 | 5.4352 |

Figure 18. Overall PnL Distribution of Basic RNN for In-Sample (Left) & Out-Of-Sample (Right) at $\alpha = 0.995$

- Risk aversion parameter ($\alpha$) = 0.997

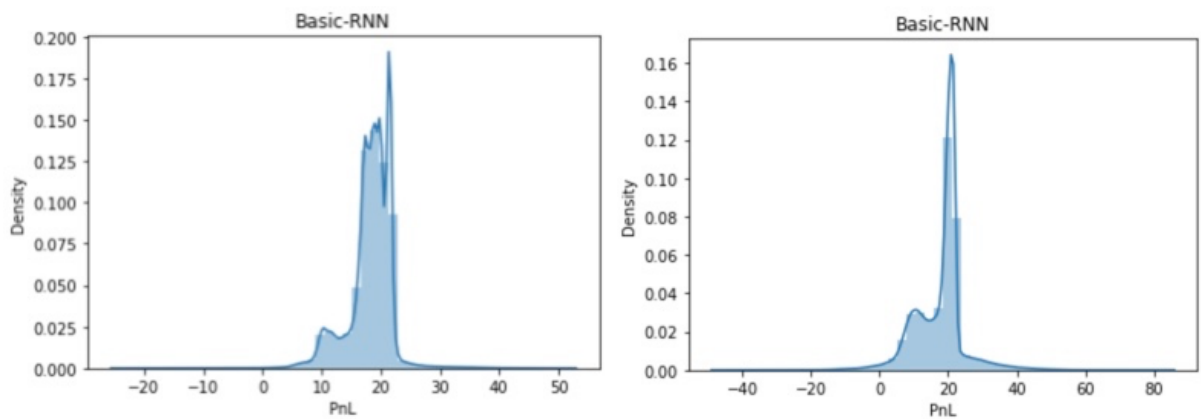

| $\alpha = 0.997$ | In-Sample | Out-of-Sample |
|---|---|---|
| Mean PnL | 18.3035 | 17.8241 |
| CVaR 1st PnL | 17.7569 | 16.4927 |
| CVaR 25th PnL | 17.0561 | 15.4533 |
| CVaR 50th PnL | 15.7375 | 12.9145 |
| CVaR 75th PnL | 13.5355 | 8.3200 |
| CVaR 99th PnL | 4.9409 | -7.8162 |
| Skewness | -0.5120 | -0.0842 |
| Excess Kurtosis | 4.7967 | 3.5754 |

Figure 19. Overall PnL Distribution of Basic RNN for In-Sample (Left) & Out-Of-Sample (Right) at $\alpha = 0.997$

- Density comparison both for in-sample and out-of-sample

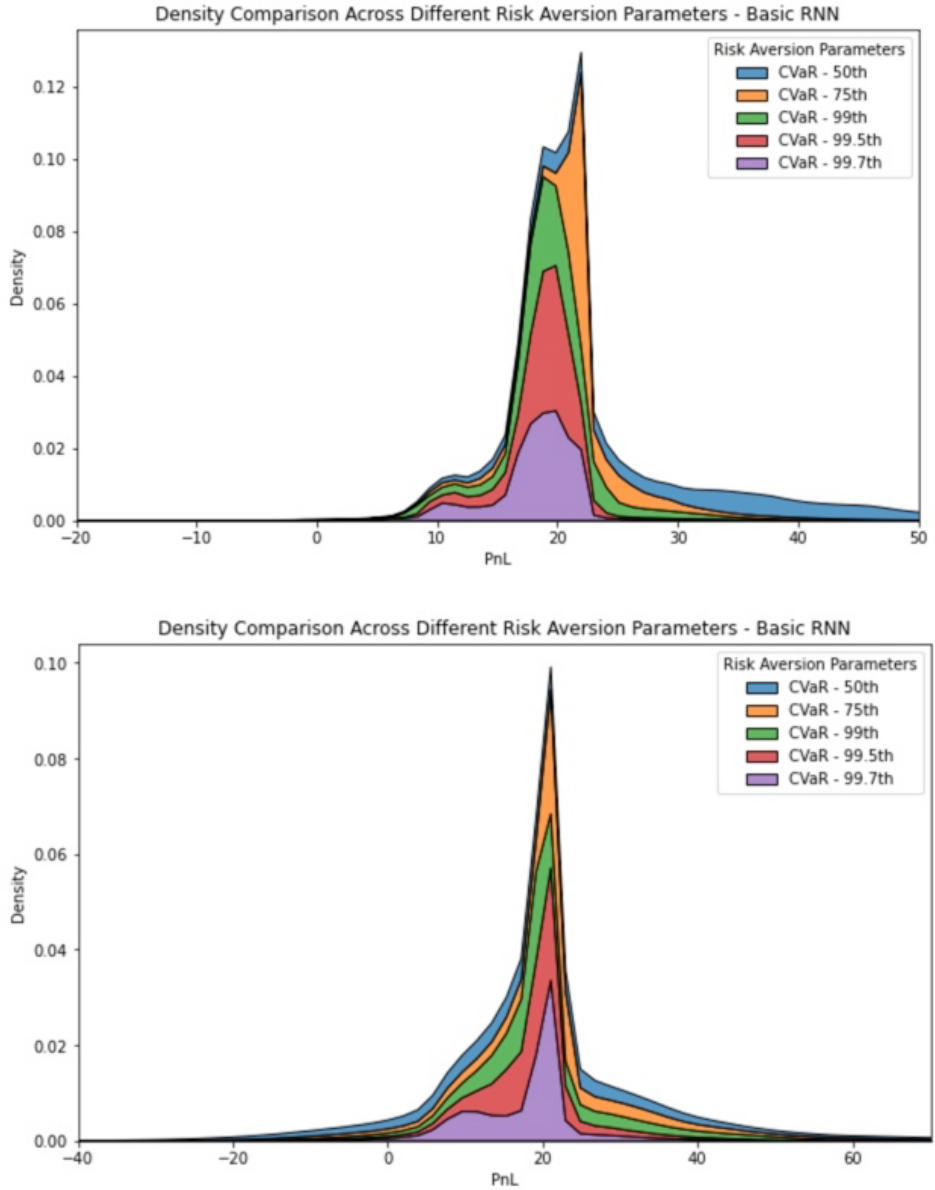

Figure 20. Density Comparison of Basic RNN for In-Sample (First) & Out-Of-Sample (Second)

- Boxplot comparison both for in-sample and out-of-sample

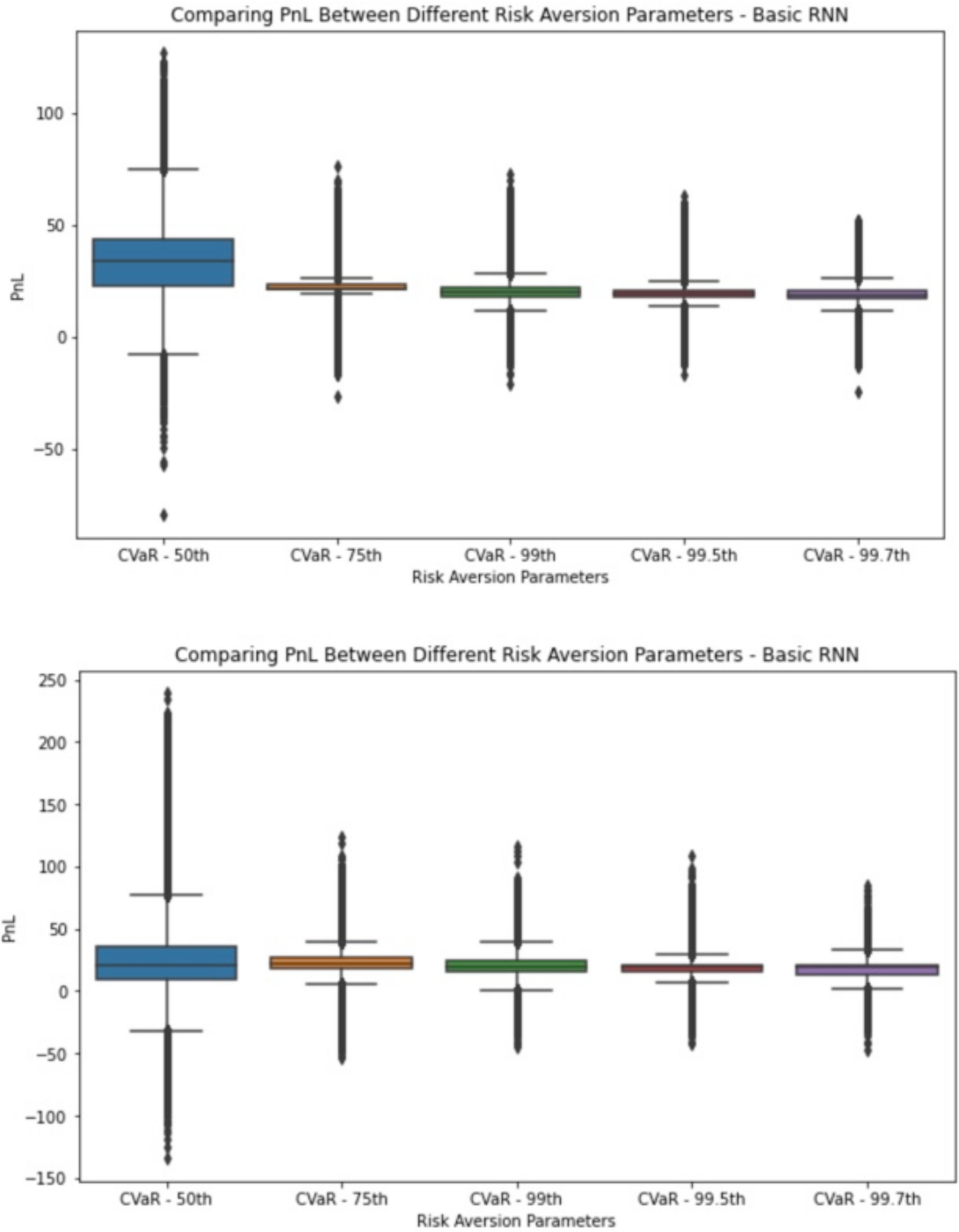

Figure 21. Boxplot Comparison of Basic RNN for In-Sample (First) & Out-Of-Sample (Second)

Interpretation

The general pattern of the original data set is captured reasonably well by synthetic data generated by time-series GANs for both training and test sets – figure 14. But, both two plots in the figure indicate there are some data points in certain sub-space that the synthetic data cannot properly capture well. Further work is needed to investigate what causes the network's inability to properly generate rather accurate synthetic data in certain sub-space (i.e., tail events – imbalance data in nature) – more detail in the directions for future work section below.

Looking at the other figures for PnL distributions across different $\alpha$ values, all distributions entail relatively higher excess kurtosis of PnL distributions compared to our experiments in Part I. This is, we believe, largely due to the very nature of fat tail empirically observed in the real financial markets as opposed to the risk-neutrality in Q-world. More importantly, the "true" fat belly part of each distribution seems to be shifted towards the right, indicating that there is a higher likelihood of gaining higher profits in general. This may, we believe, be largely thanks to the nature of multi-variate temporal prediction embedded in nonparametric scenario generator – time-series GANs.

Interestingly enough, through figures 20 and 21, we can also rather clearly see the risk-return trade-off. For instance, we see much wider dispersions of PnL distributions at $\alpha$ value of 0.5 whereas much narrower dispersions of PnL distributions are observed at $\alpha$ value of say 0.997. This can, we believe, be attributed to the fact the real markets entail the risk-return trade-off indeed quite clearly - the more risk-averse the investors are (higher values of the risk aversion parameter), the lower the risk and the returns are, vice versa.

From a theoretical perspective, we demonstrated that time-series GANs can be served as an AI-driven market path generator, extracting condensed low latent space and preserving temporal dependency from the real market data in some high-dimensional multi-variate temporal space. From a practical perspective, our hedging results showed RNN-based RL hedging agents, with nonparametrically generated data, can potentially render higher average profits with a rather evident risk-return profile observed in the real markets.

## Conclusion

In this paper, we first presented a framework of a direct policy search reinforcement agent replicating a simple vanilla European call option and use the agent for the "model-free" delta hedging. Through this, from a theoretical perspective, we demonstrated that the neural nets as a universal approximation function – in conjunction with reinforcement learning - can play a role in solving advanced mathematical problems such as stochastic differential equation (SDE) / partial differential equation (PDE) in practice. From a practical perspective, we showed how recurrent neural network (RNN)-based RL agents directly learn an optimal hedging strategy in a rather efficient manner given the risk aversion level of the investor - minimizing a convex risk measure like CVaR and producing better results particularly in terms of reducing tail risk exposures at higher values of the risk aversion parameter (i.e., 0.99) by naively embedding the additional risk aversion "constraint" in their learning processes.

Second, in an attempt to alleviate the low-dimensional parametric process constraint - in doing so potentially bridging the gap between P-world and Q-world, we illustrated nonparametric stochastic underlying scenarios generated by time-series GANs from multi-variate temporal space and demonstrated attempted hedging results on various values of risk aversion parameters. Our results showed that we can potentially achieve higher average profits with a rather evident risk-return trade-off.

We believe that this RL-based hedging framework is a more efficient way of performing hedging in practice, addressing some of the aforementioned issues with the classic models, providing promising/intuitive hedging results, and rendering a flexible framework that can be easily paired with other AI-based models (i.e., GANs or VAE-GANs) for many other purposes (i.e., signal processing in high dimensional space and real-world market path generator).

## Directions for Future Work

When it comes to directions for future research work on this front, there are many parts to be done to further improve performance and substantiate additional key considerations. We will lay out some of those both for the nonparametric stochastic generator and for the RL hedging agents.

### *Recommended Directions for Time-series GANs*

Particular attention needs to be paid to the loss construction of the embedding/reconstruction process. The current framework minimizes the classic mean square error between the original data and the reconstructed data – as in the classic auto-encoder architecture. This is not the best way of dealing with the imbalanced nature of financial data. Different loss functions better at handling extreme events (i.e., importance sampling-based weights or Q-like loss in the context of volatility proxy [5]) need to be further investigated and empirically tested with a prediction-based performance metric.

Also, the classic auto-encoder embedding structure is known to not lead to particularly useful or nicely structured latent spaces without much good compression. By overlaying Bayesian inference, variational auto-encoder (VAE) can force the model to learn rather continuous and highly structured latent space. We believe that exploring time-series VAE-GANs will be a promising research area on this front, taking into account the advantages of both VAE and GANs in generating highly accurate synthetic data and even predicting sequential data in multivariate space.

Deep Hedging, Generative Adversarial Networks, and Beyond                                      36<p style="text-align:center"><em><u>Recommended Directions for RL Agents</u></em></p>

The recurrent neural networks used for the direct policy search throughout this paper can serve as baseline neural net architectures. However, further hyper-parameter tuning and different architectural exploration (i.e., 1-D convolutional neural networks and some other non-sequential architectures) are recommended.

Also, there are a couple of key advanced techniques we can apply concerning the direct policy search RL agent [6]. One particular consideration is dealing with potential high variance in gradients by normalizing the policy gradients by overlaying some constant baseline or state-value function (i.e., actor-critic RL setup – both A2C and A3C). Also, to further encourage exploration, we can overlay an additional term in the loss function called the entropy bonus. This term will essentially indicate how much the agent is uncertain about a given action taken. If all actions have the equal probability – uniform or near-uniform, then the following entropy term produces the maximum value whereas this results in the minimum value if the policy has probability 1 of some actions and 0 for the rest of actions.

$$H(\pi) = -\sum \pi(a|s)\log(\pi(a|s)) \qquad \text{Eq. 6}$$

# References

Attilio Meucci, "'P' Versus 'Q': Differences and Commonalities between the Two Areas of Quantitative Finance," *GARP Risk Professional* (2011).

Hans Buehler et al., "Deep Hedging," arXiv preprint arXiv: 1802.03042 (2018).

Jinsung Yoon et al., "Time-series Generative Adversarial Networks," *Advances in Neural Information Processing Systems 32* (2019).

Sutton, R. S., & Barto, A. G. *Reinforcement learning: An introduction*. Massachusetts: The MIT Press, 2018.

Andrew J. Patton, "Volatility Forecast Comparison Using Imperfect Volatility Proxies," *Journals of Econometrics* (2011).

Maxim Lapan (2020). *Deep Reinforcement Learning Hands-On*. Birmingham: The Packt Publishing, 2020.